\documentstyle[aps,prl,twocolumn,epsfig]{revtex}
\begin{document}
\draft
\flushbottom
\twocolumn[
\hsize\textwidth\columnwidth\hsize\csname @twocolumnfalse\endcsname

\title{Charge Density Wave Formation in the Low-Temperature-Tetragonal Phase 
of Cuprates}
\author{N. Hasselmann$^{1}$, A.~H.~Castro Neto$^{2}$, and C.~Morais Smith$^{3}$}
\address{$^1$ Max-Planck-Institut f\"ur Physik komplexer Systeme, N\"othnitzer Str. 38, D-01187, Dresden, 
Germany \\
$^2$ Department of Physics, Boston University, 590 Commonwealth Ave, Boston, MA, 02215 \\
$^3$ Institut de Physique Th\'eorique, P\'erolles, CH-1700 Fribourg, Switzerland.}
\date{\today}
\maketitle
\tightenlines
\widetext
\advance\leftskip by 57pt
\advance\rightskip by 57pt

\begin{abstract}
We calculate the influence of transverse fluctuations on the longitudinal dynamics 
in the striped phase of cuprates by using the bosonization technique. We find 
that a charge density wave instability can arise if the stripe is quarter filled 
and the underlying lattice potential has a zigzag symmetry. Our results explain
why static stripes are experimentally observed in underdoped 
La$_{2-x-y}$Nd$_y$Sr$_x$CuO$_4$ exactly at the onset of the low-temperature-tetragonal
transition. 
\end{abstract}
]
\narrowtext

During the last years, a large number of experiments have confirmed the formation
of 1D charge stripes in doped Mott insulators \cite{mang,nick,cupr,Tran,Ichi}. 
Recently, it was shown \cite{Tran} that for La$_{2-x}$Sr$_x$CuO$_4$ co-doped with 
Nd the stripe
structure becomes static and incommensurate peaks corresponding to charge- and 
spin-stripe order can be measured by quasi-elastic neutron scattering. Moreover, it was
observed that for underdoped compounds ($x = 0.10$ and $x = 0.12$) 
the temperature $T_{co}$ at which charge
order sets in coincides with the onset of a structural transition from a 
low-temperature-orthorhombic (LTO) to a low-temperature-tetragonal (LTT) phase, 
$T_{LTT}$, whereas for the optimally doped compound $T_{co} < T_{LTT}$ \cite{Ichi}. 
Therefore, although the formation of
static stripes seems to be undeniably connected to the structural transition, it is
not yet clear {\it how} and {\it why} this should happen, and moreover, why there is
no correlation between $T_{co}$ and $T_{LTT}$ at optimal doping. 

In this letter, we address these questions and show that a charge density wave (CDW) 
instability can arise if two conditions apply simultaneously, namely, if the stripe is 
quarter filled and if the underlying lattice potential has a zigzag symmetry. This is
the situation for underdoped cuprates in the LTT phase: neutron 
scattering experiments have shown that for $(0.06 < x < 0.12)$ the stripes are
incompressible, with one hole every two lattice sites \cite{cupr,Tran}. In addition, 
within the LTT phase the buckling of the oxygen octahedra has 
the zigzag symmetry needed to stabilize bond centered stripes. The consequence of 
the combined presence of both facts leads to the formation of static stripes. 
 
We calculate the influence of transverse fluctuations on the
longitudinal dynamics along the stripes. The longitudinal modes are described as 
Luttinger liquids. The coupling between the
longitudinal and transverse modes arises because the electrons moving along the
stripe are scattered from transverse kinks. By writing the Hamiltonian in
a bosonic representation, we find several forward scattering channels 
and one $4 k_F$ scattering momentum transfer from
longitudinal to transverse modes. The latter becomes relevant at quarter filling and
leads to the formation of a longitudinal CDW if the
transverse modes are frozen with a zigzag symmetry.

Before introducing our model of a Luttinger liquid
coupled to transverse kinks, we briefly review the transverse dynamics of
a chain of holes. This problem was considered
by Zaanen et al. \cite{Zaan}, who showed its equivalence to a quantum spin-1 chain.
The full phase diagram for this problem was determined numerically by den Nijs and 
Rommelse \cite{denN} after earlier calculations by Schulz \cite{Schu}, who treated the 
spin-1 problem as two coupled spin-1/2 chains. Here, we review Schulz calculations 
\cite{Schu} and generate the full phase diagram from this formalism. Then, we study the
effect of a coupling between the longitudinal and the transverse sector to determine 
when the operator responsible for the CDW instability becomes relevant.
 
Our starting point for the transverse modes is the spin-1 Hamiltonian \cite{Zaan}
$$
H = \sum_n \left[ - t \left( S_n^x S_{n+1}^x +S_n^y S_{n+1}^y
\right) - J_z S_n^z S_{n+1}^z + d (S_n^z)^2 \right].
$$
By replacing the spin-1 operator by a sum of two spin 1/2 operators, 
${\bf S}_n=\mbox{\boldmath $\tau$}_{a}(n) + \mbox{\boldmath $\tau$}_{b}(n)$, 
performing a Jordan-Wigner transformation into a system of interacting
fermions, and then using a standard bosonization technique, one finally obtains 
$H = H_+ + H_-$ with \cite{Schu}
\begin{eqnarray}
\label{transcont}
H_+&=& \frac{u_+}{2 \pi} \int dx \left[ K_+ \pi^2 \chi_+^2 +
K_+^{-1} \left(\partial_x \Psi_+ \right)^2 \right] \\ \nonumber 
&+&  
\frac{\mu_1}{\pi^2 \alpha^2}
\int dx  \cos \left[\sqrt{8} \Psi_+ \right] ;  \\ \nonumber
H_-&=&  \frac{u_-}{2 \pi} \int dx \left[ K_- \pi^2 \chi_-^2 +
K_-^{-1} \left(\partial_x \Psi_- \right)^2 \right] \\ \nonumber 
&+& \frac{\mu_2}{\pi^2 \alpha^2}
\int dx  \cos \left[\sqrt{8} \Psi_- \right] 
+ \frac{\mu_3}{\pi^2 \alpha^2}
\int dx  \cos \left[\sqrt{2} \Theta_- \right].
\end{eqnarray}
Here, $\Psi_\pm=(\Psi_a \pm \Psi_b)/\sqrt{2}$ and 
$\chi_\pm=(\chi_a \pm \chi_b)/ \sqrt{2}$. The bosonic phase fields are  
\begin{eqnarray} \nonumber 
\Psi_{a/b} (x) &=& -\frac{i \pi}{\ell }\sum_{k\neq 0} \frac{1}{k}
\left[\tilde{\rho}_{R,a/b}+\tilde{\rho}_{L,a/b} \right]
e^{-\alpha |k|/2 - i k x}, \\ 
\chi_{a/b} (x) &=& \frac{1}{\ell } \sum_{k\neq 0} 
\left[\tilde{\rho}_{R,a/b}-\tilde{\rho}_{L,a/b} \right]
e^{-\alpha |k|/2 - i k x},
\end{eqnarray}
where $\alpha$ is the lattice cutoff and $\ell$ the system size. 
The fields with different $a/b$ indices commute and 
$\left[\Psi_{a/b}(x),\chi_{a/b}(y)\right]=i \delta(x-y)$. 
The $\Psi_{a/b}$
are related to density fluctuations of the $a$,$b$ fermions through 
$\partial_x \Psi_{a/b}(x)=- \pi [\tilde{\rho}_{R,a/b}(x) +
\tilde{\rho}_{L,a/b}(x)- 1 /2 ].$ 
The velocities of the acoustic excitations are 
$u_+ \simeq t \sqrt{1+2 (d-3 J_z)/(\pi t)}$ 
and $u_- \simeq t \sqrt{1-2(d+J_z)/(\pi t)}$. Further,
$K_+= [1+2 (d-3 J_z)/(\pi t)]^{-1/2}$, $K_- = [1-2(d+J_z)/(\pi t)]^{-1/2}$
and $\Theta_{-}(x)=\pi \int^x dx^{\prime} \chi_{-}(x^{\prime})$.
The coupling constants for the nonlinear terms are approximately given by 
$\mu_1=\mu_2=J_z+ d$ and $\mu_3=-t$. 
While the $\mu_1$ and $\mu_2$ terms 
represent, respectively, back- and forward scattering
contributions, the $\mu_3$ term originates 
from a non-local Fermi coupling and is the so-called disorder operator. 
The most relevant operator resulting from the original Hamiltonian is
$\cos \left(\sqrt{2}\Theta_- \right)$. There are other operators
which mix the ($+$) and ($-$) sectors but their scaling dimension is
such that they are never dominant.

To analyze the possible phases of the model defined through
Eq.~(\ref{transcont}), 
we treat the non-linear terms as perturbations and consider their
scaling behavior \cite{kadanoff79}.
The $\mu_1$ term is relevant
for $d>3 J_z$. The transition at $d=3 J_z$ is the roughening
transition of the string, which can be seen calculating the large
$n$ behavior of the correlator of the transverse displacements
$G_u(n) = \left< u(0)u(n)\right>=
\sum_{i,j=0}^{n-1} \left<S_i^z S_j^z \right>$, where $S^z_i 
= u_{i + 1} - u_i$.
Using the  bosonized form 
$
S^z(x) \simeq - (\sqrt{2} / \pi) \partial_x \Psi_+ + (2/ \pi \alpha)
e^{i \pi x} \cos(\sqrt{2} \Psi_+) \cos(\sqrt{2} \Psi_-), 
$
we find the equal time correlation function
$ G_z = \left< S^z(0) S^z(x) \right> \simeq ( 2K_+ / \pi^2 x^2) + C_z e^{i \pi x} 
\left|x\right|^{-2K_+ -2 K_-} $
which consists of a smooth and an oscillating part
($C_z$ is a constant). The smooth part
gives rise to a logarithmic divergence of $G_u(n)$ for
$K_+\neq 0$. However, 
for $d>3 J_z$ the $\mu_1$ operator is relevant and $K_+$ scales
to zero, killing the logarithmic divergence of $G_u(n)$. 

The ($-$) sector is more involved. The
$\mu_2$ operator is relevant for $d+J_z<0$ and  $\mu_3$ 
is relevant for $d+J_z>-15 \pi t/2$ \cite{Schu}. 
Thus, the gaussian fixed point
is never stable and the ($-$) sector always flows to
strong coupling. 
This implies that a perturbative approach starting from
the Gaussian model cannot be trusted. However, 
it was pointed out
by den Nijs \cite{denN,Schu} that the ($-$) sector has an Ising symmetry.
It has a form identical to the continuum limit of a classical 2D  XY  model
$H_{xy}=\sum_{\left< ij\right>}\cos(\phi_i - \phi_j)$ with a 
2-fold (Ising) symmetry breaking term of the form $\cos 2 \phi_i$. We 
therefore expect a transition in the ($-$) sector of the Ising type.
We locate the critical line of the transition as the
line where the scaling dimensions of the two non-linear operators
in $H_-$ are equal, $d+J_z\sim -3 t \pi/2$. 

To analyze the properties of the phases we investigate the behavior
of the equal time correlation function 
$G_{\perp}(n)=\left< S^+(n) S^-(0)\right>$. 
If $\Psi_+$ is ordered (i.e. $d>3 J_z$), then 
$G_{\perp}$ decays exponentially as $\Theta_+$ correlations are
short ranged. Similarly, if $d<3 J_z$ and
$d+J_z> -3 t \pi/2$, such that $\Psi_-$ has long range order, then
$\Theta_-$ correlations also are short ranged, again leading
to an exponentially $G_{\perp}$. However, for $d<3 J_z$
and $d+J_z< -3 t \pi/2$, $\Theta_-$ is ordered and $G_{\perp}$
decays algebraically, $ G_\perp (x)\sim |x|^{-1/4K_+}$. 
Combining the results from the ($+$) and ($-$) sectors, we then
find five different phases. 
1) a gap-full flat phase, with dominant $\mu_1$ and 
$\mu_3$ perturbations, exponentially decaying $G_z$ and $G_\perp$ 
correlations and a finite limit of $G_u$. 2) A gap-less rough phase in which
only $\mu_3$ is relevant, with an algebraic decay in the smooth part
of $G_z$, exponentially decaying  $G_\perp$ and logarithmically
divergent $G_u$ correlations. 3) A gap-less bond centered (BC) rough
phase with a zig-zag pattern (anti-ferromagnetic correlations with 
$S_z=\pm 1$ but no $S_z=0$ states). This phase differs from the rough phase
in that it has an algebraic decay in the $G_\perp$ correlations.
4) A BC flat phase with
dominant $\mu_2$ and $\mu_3$ perturbations. Like the BC rough phase,
this phase has a zig-zag pattern which now is however long ranged. 
This phase is gaped.
5) For $3 J_z -d > \pi t /2$, the present analysis cannot be
applied because the diagonalization of the quartic $a,b$ interactions
through a Bogoliubov transformation breaks down, leading to an 
unphysical purely imaginary value of $K_+$. As argued by Schulz
\cite{Schu}, this signals a transition to a ferromagnetic state, i.e. a diagonal
stripe state. 

Besides the five phases found by Schulz, den Nijs \cite{denN} has identified a sixth 
phase, see Fig.\ 1. The additional phase
found by den Nijs is a disordered flat phase (DOF), which
is gap-full. This phase has, in contrast to the flat phase, a finite density
of kinks and anti-kinks ($S_z=\pm 1$ states) but the $G_u$ correlator
does not show a logarithmic divergence, making this phase different
from the rough phase. In the DOF phase the kinks are positionally disordered,
but have an anti-ferromagnet order in the sense that 
a kink $S_z=1$ is more likely to be followed by an anti-kink $S_z=-1$
rather than another kink, with any number of $S_z=0$ states in between
them. In spin language, the DOF phase is 
the valence bond phase which is responsible for the Haldane gap.

In fact, there are signs of this transition also in the Abelian
bosonization approach. At $d+J_z=0$, $\mu_1$ and
$\mu_2$ changes sign. This is unimportant for $\mu_2$ as
in this parameter regime the ($-$) sector is dominated by the
$\mu_3$ and not the $\mu_2$  operator. However, if the $\mu_1$
operator is relevant, then a sign change of $\mu_1$ is of
consequence. Typically, such a sign change is indicative of
a competition between two fixed points. A similar situation
occurs in a spin-$1/2$ chain with nearest and next-nearest
interactions, where dimerized and anti-ferromagnetic ground states
compete \cite{Hald}. In the present situation, the
competition is between on-site ($d$) and nearest neighbor interaction
($J_z$).
A useful order parameter to distinguish between the flat and
the disorder flat phases is the parity of the steps, $P =
\left< \cos (\pi u_n) \right>$, which in bosonized form becomes
$P\sim \left< \cos \left[ \sqrt{2}\left(\Psi_+(x)-\Psi_+(0) \right) \right] \right>$, 
with $x=0$
being the boundary of the string. If the $\mu_1$ operator is relevant,
the field $\Psi_+$ is pinned and a sign change in $\mu_1$ leads
to a $\pi$ phase shift of $\sqrt{8} \Psi_+$. As a result, $P$ shifts
from zero to a finite value (if the boundary term is left unchanged).
Therefore, while the Abelian bosonization approach does not allow for a
detailed study of the preroughening transition or the DOF phase,
the existence of this transition can be readily inferred.
\begin{figure}
\hspace{1.5cm}
\epsfig{file=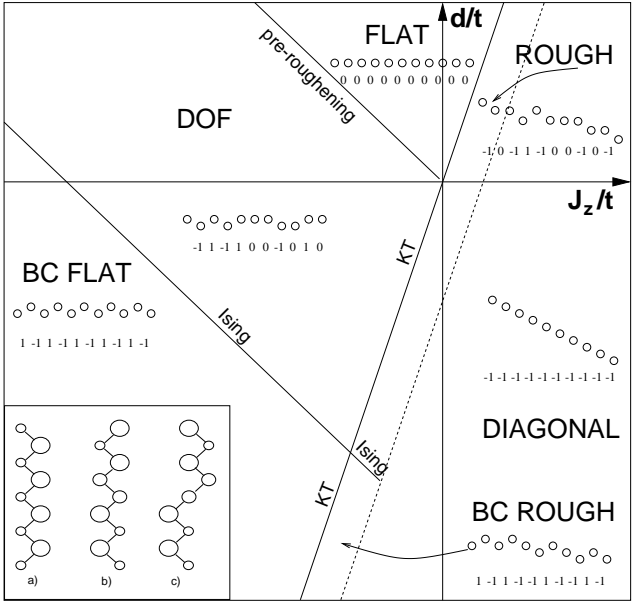,height=8cm}
\vspace{0.3cm}
\caption[]{\label{zigzagpic}Phase diagram for the spin-1 chain
and the corresponding stripe configurations. The inset in the
left corner shows a 4 $k_F$ CDW state in the BC flat phase.
The size of the circle represents the local density of holes.
a) commensurate ground state. b) CDW domain wall soliton. c) 
soliton state of both the transverse and longitudinal order.}
\end{figure}
We now study a model including both, transverse
and longitudinal modes.
The longitudinal modes along the stripe are described by a Luttinger liquid \cite{Emery},
\begin{eqnarray}
\label{1dham} \nonumber
H_\rho &=& \frac{u_\rho}{2 \pi} \int dx \left[ K_\rho \pi^2 \Pi_\rho^2 +
K_\rho^{-1} \left(\partial_x \Phi_\rho \right)^2 \right]
;  \\ \nonumber 
H_\sigma&=&  \frac{u_\sigma}{2 \pi} \int dx \left[ K_\sigma \pi^2 \Pi_\sigma^2 
+ K_\sigma^{-1} \left(\partial_x \Phi_\sigma \right)^2 \right] \\ 
&+& \frac{g_1}{2 \pi^2 \alpha^2}
\int dx  \cos \left[\sqrt{8} \Phi_\sigma \right],
\end{eqnarray}
where $g_1$ is the backward scattering strength. 
The fields $\Phi_{\rho/\sigma}=(\Phi_\uparrow\pm
\Phi_\downarrow)/\sqrt{2}$, $\Pi_{\rho/\sigma}=(\Pi_\uparrow\pm
\Pi_\downarrow)/\sqrt{2}$ are related to the Fermi fields
of left- and right-moving particles.
The parameters of the bosonic model depend on the underlying
lattice model. We assume repulsive interactions, therefore $K_\rho<1$ and 
$K_\sigma>1$.
The backscattering term is irrelevant under the renormalization group procedure 
and will be neglected here \cite{Emery}.

Physically, the coupling of the 
transverse and longitudinal modes arises from
the scattering of the electrons which move along
the stripe from transverse kinks. This interaction
is local and we can express it in terms of the longitudinal
electron density $\rho_{\uparrow/\downarrow}$ and the
kink density $S_z^2$. We will consider only terms that couple
the charge density to the kink density. Terms that couple
the kinks and the spin density are irrelevant in presence
of repulsive interactions and we thus omit them. 
A local coupling of the electron density to
the kink density can thus be written as
\begin{eqnarray}
H_{c1} &= & \gamma_1 \int dx \left( \rho_\uparrow + \rho_\downarrow \right)
\left( S^z \right)^2 \nonumber  \\
H_{c2} &= & \gamma_2 \int dx \rho_\uparrow  \rho_\downarrow 
\left( S^z \right)^2 
\end{eqnarray}
with coupling constants $\gamma_1$,$\gamma_2$.
These terms can be readily bosonized. From $H_{c1}$ one
obtains the following terms,
\begin{eqnarray}  \nonumber
H_{c1}  \sim &&\int dx \ \partial_x \Phi_\rho \left[ \frac{2 \gamma_1}{\pi^2} \ 
\partial_x \Psi_+  + \lambda_1  \cos \left( \sqrt{8} 
\Psi_- \right) \right. \\ 
&& \left. + \lambda_2  \cos \left( \sqrt{8} \Psi_+ \right) \right] 
+ \mbox{irrelevant terms},
\end{eqnarray}
with 
$\lambda_i = \gamma_i / \left( \sqrt{2} \pi^3 \alpha^2\right)$.
The first term represents forward scattering 
between the acoustic transverse and 
longitudinal charge modes. In general, these two modes
have different velocities $u_+\neq u_\rho$ and thus
this interaction is retarded as can be seen by 
performing a Gaussian average over one of the modes. 
This first term and the Gaussian
parts of $H_+$ and $H_\rho$  and
can be jointly diagonalized, leading to a hybridization of the transverse
and longitudinal modes. This leads to an instability 
of the system which is similar to the Wentzel-Bardeen instability
in one dimensional metals coupled to phonons.
However, this instability occurs only at a very large coupling 
$\gamma_1\sim 4\sqrt{\left(\pi K_+ K_\rho\right)/ \left( u_\rho u_+ \right)}$
\cite{loss94}. For small $\gamma_1$, the case considered here, the
zero momentum transfer interaction is not very efficient \cite{voit90}. Also,
as the hybridization is small for small $\gamma_1$, the scaling analysis
below is only weakly affected by the hybridization
and we thus ignore this correction.
The $\lambda_1$ and $\lambda_2$ terms are only important if
the respective cosine terms have a finite expectation value.
If they do, these terms act like a shift of the chemical potential
of the electrons, as $\partial_x \Phi_\rho$ measures the deviation from
the charge density of its equilibrium value.
These terms compete however with terms generated by
$\gamma_2$, as will be discussed below.
Bosonization of the $\gamma_2$ interaction gives the following terms,
\begin{eqnarray} \nonumber 
&& H_{c2} \sim  \int dx \ \left[ \lambda_3 \  \partial_x \Psi_+
\cos \left(\sqrt{8} \Phi_\sigma \right) + 
  \lambda_4  \cos \left(\sqrt{8} \Phi_\sigma
\right) \times \right. \\ &&  \cos \left( \sqrt{8} \Psi_+ \right)
+ \lambda_5   \cos \left(\sqrt{8} \Phi_\rho
-\sqrt{2} \Psi_+ + \left(4 k_F - \pi \right) x
\right) \times \nonumber  \\ && \left. 
\cos \left( \sqrt{2} \Psi_- \right) \right]   
+ \ \ \mbox{irrelevant terms}
\end{eqnarray}
The $\lambda_3$ and $\lambda_4$ terms result from scattering involving
two $a$,$b$ operators whereas $\lambda_5$ involves four $a$,$b$ operators.
While the first two terms describe forward scattering, 
the $\lambda_5$ term results from processes with momentum transfer 
$4 k_F$ from longitudinal to transverse modes. At $k_F=\pi/4$,
i.e. at quarter filling of the stripe this term is important, as 
the oscillatory $x$ dependence vanishes.  

To understand the effect of the longitudinal-transverse coupling
we examine the scaling dimensions of the $\lambda_i$ operators.
We envision strongly repulsive interactions among the
electrons on the stripe with $K_\rho<1$ and $K_\sigma>1$. The
$\lambda_4$ operator is relevant only for $K_\sigma + K_+ < 1$ and thus
unimportant. Similarly, $\lambda_3$ can be neglected as it is relevant
only for $K_\sigma<1/2$.
The $\lambda_1$ operator is relevant for $K_-<1/2$, 
$\lambda_2$ is relevant for $K_+<1/2$ and 
the $\lambda_5$ operator
is relevant for $2 K_\rho + K_+ + K_-<2$. 

Let us consider that $K_+<1$, i.e. the ($+$) sector
is massive. In that case, the transverse modes behave at large wavelengths
as if $K_+=0$. Then, if $2 K_\rho + K_- < 2$, the $\lambda_5$ operator
will become relevant at large scales. The $\lambda_5$ term pins the 
longitudinal charge modes to the transverse modes and induces a longitudinal
CDW. This is seen for the case with $K_-<1/2$,
i.e. in the BC flat phase, where both $\Psi_+$ and $\Psi_-$ have  long
range order. 
The freezing of the transverse modes
in a zig-zag pattern leads to a $\pi$ periodic potential for the
transverse modes and allows for Umklapp scattering of the quarter
filled stripe which causes a longitudinal $4 k_F$ CDW.
Thus, a relevant $\lambda_5$ term makes the $\Phi_\rho$
field massive and pins it to the frozen kink-antiking pattern
of the transverse fluctuations. Such a stripe with a frozen kink pattern
and a longitudinal CDW pinned to it is depicted in Fig.~\ref{zigzagpic}a.
If the $\lambda_5$ term is relevant, it is
in direct competition with the $\lambda_{1,2}$
terms which favor a non-zero $\partial_x \Phi_\rho$. While
an exact treatment of this competition is not currently possible, we can gain
some insight if we assume mean field values for the transverse fields.
Deep into the BC flat phase we may replace $\cos \left(\sqrt{8} \Psi_\pm
\right)$ in the $\lambda_{1,2}$ terms by $\left<\cos \left(\sqrt{8} \Psi_\pm
\right)\right>$. Similarly, the $\lambda_5$ term becomes % may be written
$ \lambda_5 \int dx \ \cos \left(\sqrt{8} \Phi_\rho \right)
\left<\cos \left(\sqrt{2} \Psi_+
\right)\right> \left<\cos \left(\sqrt{2} \Psi_-
\right)\right> $.
In this approximation the longitudinal charge sector has a Hamiltonian
which is well known from studies of commensurate-incommensurate transitions
\cite{schulz80,pokrovsky79}. Such a model can be studied within 
a fermion description in terms of spin-less holons 
\cite{schulz80,pokrovsky79}, in which the $\lambda_5$ Umklapp term
becomes quadratic rather than quartic as in the original Fermi
operators. In these models, a gaped commensurate state, which exists
for small $\lambda_{1,2}$ (i.e. a small prefactor of the $\partial_x
\Psi_\rho$ term) is followed by an incommensurate state at large
values of $\lambda_{1,2}$ where the gap is destroyed and a small
number of holons move as weakly interacting particles. 
These holons are solitonic in character and can be interpreted
within the present framework as point like domain ``walls'' of the
$\pi$ periodic CDW. Such a CDW soliton state, which
involves only the transverse sector, is depicted in 
Fig.~\ref{zigzagpic}b. A different soliton is also possible, 
see Fig.~\ref{zigzagpic}c, which represents a domain wall for both, transverse 
and longitudinal degrees of freedom.

It is interesting, that even if $K_-> 1/2$, i.e. in the DOF
phase, the $\lambda_5$ term can
be relevant for sufficiently strong repulsion (small
$K_\rho$). However,
if one assumes a Hubbard model for the longitudinal
modes, the lower limit for $K_\rho$ is $1/2$ so that in the flat
phase, with $K_{-}>1$, the $\lambda_5$ operator is always irrelevant.
This result has two important consequences: first, since $\lambda_5$
is irrelevant in the flat phase, any kind of underlying structure 
which favors site-centered stripes is unable to produce a CDW, implying
that the usual stripe ``cartoon'' \cite{Tran} is actually
misleading. Second, it allows us to understand why a static CDW is formed
within the LTT phase of under-doped cuprates, since the symmetry of this 
phase favors a bond-centered pattern, for which the $\lambda_5$ operator
is relevant. 

In conclusion, we evaluated the effect of transverse fluctuations
on the longitudinal transport in a Luttinger liquid stripe phase. 
We found that a CDW instability arises if the stripes are quarter 
filled and the underlying potential has a zigzag symmetry. Our results
explain why static CDWs are formed in under-doped cuprates exactly at
the onset of the LTT structural transition.

C.~M.~S. was supported by Swiss National Foundation under grant 
620-62868.00.

\end{document}